\documentclass[12pt,fleqn]{article}
\usepackage[DIV12]{typearea}
\usepackage{amsmath,amsfonts,amssymb}
\usepackage{graphicx}
\usepackage{cite}
\usepackage{mathrsfs}
\usepackage{bbm}
\usepackage{pifont}
\usepackage{marvosym}
\usepackage{units}
\usepackage[small,loose]{subfigure}  
\usepackage{xcolor}

\usepackage{hyperref}

\newcommand{\Figref}[1]{figure~\ref{#1}}
\newcommand{\Tabref}[1]{table~\ref{#1}}
\newcommand{\Secref}[1]{section~\ref{#1}}

\newcommand{\eVdist}{\kern-0.06em}

\newcommand{\GeV}{\text{Ge\eVdist V}}

\DeclareMathOperator{\re}{Re}
\DeclareMathOperator{\im}{Im}

\newcommand{\D}{\mathrm{d}}
\newcommand{\I}{\mathrm{i}}

\newcommand{\SO}[1]{\ensuremath{\mathrm{SO}(#1)}}
\newcommand{\SU}[1]{\ensuremath{\mathrm{SU}(#1)}}
\newcommand{\U}[1]{\ensuremath{\mathrm{U}(#1)}}
\newcommand{\Z}[1]{\ensuremath{\mathbbm{Z}_{#1}}} 

\newcommand{\hu}{\ensuremath{H_u}}
\newcommand{\hd}{\ensuremath{H_d}}
\newcommand{\qhu}{\ensuremath{q_{\hu}}}
\newcommand{\qhd}{\ensuremath{q_{\hd}}}
\newcommand{\dilaton}{\ensuremath{S}}
\newcommand{\MP}{\ensuremath{M_\mathrm{P}}}

\newcommand{\rep}[1]{\ensuremath{\boldsymbol{#1}}}
\newcommand{\crep}[1]{\ensuremath{\overline{\boldsymbol{#1}}}}

\allowdisplaybreaks[1]

\numberwithin{equation}{section}
\numberwithin{table}{section}

\def\BLAU{\empty}

\def\mytitle{Supersymmetric unification and $\boldsymbol{R}$ symmetries}
\title{\mytitle}

\begin{document}

\begin{titlepage}

\begin{flushright}
 UCI-TR-2012-19\\
 TUM-HEP 869/12\\
 FLAVOR-EU-32\\
 CETUP-016\\
\end{flushright}

\vspace*{1.0cm}

\renewcommand*{\thefootnote}{\fnsymbol{footnote}}
\begin{center}
{\Large\bf
\mytitle\footnote[1]{Based on invited talks given at 
\href{http://www.newton.ac.uk/programmes/BSM/bsmw05.html}{String Phenomenology
2012}, 
\href{http://www.dsu.edu/research/cetup/general.html}{CETUP 2012}, 
\href{http://pacific.physics.ucla.edu/}{PACIFIC 2012},
\href{https://indico.desy.de/conferenceDisplay.py?confId=5694}{4th Bethe Center
Workshop} and the
\href{http://intern.universe-cluster.de/indico/event/2751}{Universe Cluster
Science Week 2012}.}
}
\renewcommand*{\thefootnote}{\arabic{footnote}}

\vspace{1cm}

\textbf{Mu--Chun Chen\footnote[1]{Email: \texttt{muchunc@uci.edu}}{}}
\\[3mm]
\textit{\small
Department of Physics and Astronomy, University of California,\\
~~Irvine, California 92697--4575, USA
}
\\[5mm]
\textbf{
Maximilian Fallbacher\footnote[2]{Email:
\texttt{maximilian.fallbacher@ph.tum.de}}{} and 
Michael Ratz\footnote[3]{Email: \texttt{michael.ratz@tum.de}}{}
}
\\[3mm]
\textit{\small
Physik Department T30, Technische Universit\"at M\"unchen, \\
~~James--Franck--Stra\ss e, 85748 Garching, Germany
}
\end{center}

\vspace{1cm}

\begin{abstract}
We review the role of $R$ symmetries in models of supersymmetric unification in
four and more dimensions, and in string theory.  We show that, if one demands
anomaly freedom and fermion masses, only $R$ symmetries can forbid the
supersymmetric Higgs mass term $\mu$. We then review the proof that $R$
symmetries are not available in conventional grand unified theories (GUTs) and
argue that this prevents natural solutions to the doublet--triplet splitting
problem in four dimensions. On the other hand, higher--dimensional GUTs do not
suffer from this problem. We briefly comment on an explicit string--derived
model in which the $\mu$ and dimension five proton decay problems are solved
\emph{simultaneously} by an order four discrete $R$ symmetry. We also
comment on the higher--dimensional origin of this symmetry. 
\end{abstract}

\end{titlepage}

\section{Introduction and outline}

The minimal supersymmetric standard model (MSSM) provides an attractive scheme
for physics beyond the standard model (SM) of particle physics.
The MSSM has the following, attractive features: 
\begin{itemize}
 \item it is based on supersymmetry, which is, under certain modest
 assumptions, the maximal extension of the Poincar\'e symmetry of our
 four--dimensional Minkowski space--time;
 \item it provides automatically a dark matter candidate, which is stable due to
 the $\Z2^\mathcal{M}$ matter parity;
 \item supersymmetry allows us to stabilize the gauge hierarchy against
radiative corrections. 
\end{itemize}
In the context of unified theories, the perhaps most important property of the MSSM
is that it provides us with the very compelling picture of precision gauge
coupling unification \cite{Dimopoulos:1981yj}. That is, if one assumes that the
superpartners have masses of the order TeV and extrapolates the gauge couplings
$g_i$ of the SM gauge factors \SU3, \SU2 and \U1 to higher energies, one finds
that they meet with a high precision at the scale of a
$\text{few}\times10^{16}\,\GeV$. This property of the MSSM represents,
given the still persisting lack of evidence for superpartners at the LHC, perhaps the
greatest motivation for supersymmetry.
Arguably, the most compelling explanation of this fact arises if the SM gauge
group is embedded in a simple gauge group, specifically 
\begin{equation}
 G_\mathrm{SM}~=~\SU3\times\SU2\times\U1~\subset~\SU5
\end{equation}
or a group containing \SU5.

This brings us to the scheme of grand unified theories (GUTs). Specifically,
GUTs based on the gauge groups \SU5 and \SO{10} have many appealing features
(for a review see, e.g., \cite{Raby:2008gh}):
\begin{enumerate}
\item GUTs explain charge quantization;
\item they simplify the matter content. The five irreducible representations
(irreps) forming one generation of SM matter can be grouped into two \SU5
irreps \cite{Georgi:1974sy},
\begin{equation}
\text{SM generation}~=~\rep{10}+ 
\crep{5}\;.
\end{equation}
A further simplification of the matter sector happens in \SO{10}
\cite{Fritzsch:1974nn}, where 
\begin{eqnarray}
 \rep{16} & = &
 \rep{10}\oplus\crep{5}\oplus\rep{1}
 \nonumber\\
 & = & 
 \text{SM generation with `right--handed' neutrino}\;.
\end{eqnarray}
\end{enumerate}
One of the main assumptions of this review is that these features are not by
accident.

In this review, we will specifically discuss the role of (discrete) $R$ symmetries in
supersymmetric models of unification. After a short review of some of the issues of the
MSSM, we will discuss the importance of anomaly constraints and in particular
``anomaly universality'' for their resolution. Using these techniques, we will
show that only $R$ symmetries can forbid the $\mu$ term in the MSSM.
Furthermore, as we will then argue, these $R$ symmetries are already almost
uniquely determined by the anomaly universality conditions. However, given
certain general assumptions which we will specify, $R$ symmetries are not
available in four--dimensional models of grand unification. On the other hand,
$R$ symmetries are available in higher--dimensional and, in particular, in
stringy settings, where they arise as discrete remnants of the Lorentz symmetry
of compact space. We will comment on explicit models where precisely the
phenomenologically desired symmetries arise this way. Finally, we will provide a
short summary.

\section{The MSSM, grand unification and all that}

We start by reviewing the problems of the MSSM in \Secref{sec:MSSMProblems} and
describe specifically the proton decay problems in \Secref{sec:ProtonDecay}. As
we shall see, the conventional solutions to the MSSM problems are,
arguably, not fully satisfactory.

\subsection{Problems of the MSSM}
\label{sec:MSSMProblems}

As is well known, the MSSM has, besides many desired features, certain
shortcomings.  Several of them are connected to the appearance of operators in
the superpotential which are consistent with all symmetries of the MSSM but have
phenomenologically undesired effects, or are plainly inconsistent with
observation. The gauge invariant superpotential terms
up to order four include
\begin{eqnarray}
\mathscr{W} & = & 
\mu\, \hd \hu + \kappa_i\, L_i \hu 
\nonumber \\
& &{}+ Y_e^{ij}\, L_i \hd \overline{E}_j + Y_d^{ij}\, Q_i \hd
\overline{D}_j + Y_u^{ij}\, Q_i \hu \overline{U}_j 
\nonumber\\
& & {} + \lambda_{ijk}\, L_i L_j \overline{E}_k 
+ \lambda^\prime _{ijk}\, L_i Q_j \overline{D}_k  
+ \lambda^{\prime\prime}_{ijk}\, \overline{U}_i\overline{D}_j\overline{D}_k 
\nonumber\\
& & {} 
+\kappa^{(0)}_{ij}\, \hu L_i\,\hu L_j
+\kappa^{(1)}_{ijk\ell}\, Q_i Q_j Q_k L_\ell 
+ \kappa^{(2)}_{ijk\ell}\,
\overline{U}_i\overline{U}_j\overline{D}_k \overline{E}_\ell\;.
\label{eq:WMSSM}
\end{eqnarray}
Here, in an obvious notation, $\hu$ and $\hd$ denote the MSSM Higgs doublets,
and $Q_i$, $\overline{U}_i$, $\overline{D}_i$, $L_i$ and $\overline{E}_i$ the
three generations of MSSM matter. The $\mu$ term in the first line, for
phenomenological reasons, has to be of order TeV, and the Yukawa couplings
$Y_e^{ij}$, $Y_u^{ij}$ and $Y_d^{ij}$ are required in order to describe fermion
masses.  Moreover, the perhaps simplest explanation of small Majorana neutrino
masses needs a non--trivial $\kappa^{(0)}_{ij}$ of the order
$(10^{14}\,\GeV)^{-1}$.

Unfortunately, there are various additional terms, which turn out to be very
problematic. First of all, the so--called $R$ parity violating couplings
$\kappa_i$, $\lambda_{ijk}$, $\lambda^{\prime}_{ijk}$ and
$\lambda^{\prime\prime}_{ijk}$  are strongly constrained by the experiments,
i.e.\ they have to be either very small or completely absent (cf.\ e.g.\
\cite{Allanach:1999ic}). Secondly, there are strong bounds on the coefficients
$\kappa^{(1,2)}_{ijk\ell}$ of the so--called  dimension--five proton decay
operators.  This shows that supersymmetry alone is not a viable theory. It has
to be amended by some additional structure, preferably by symmetries which
ensure that the phenomenological predictions of the extended model are in
agreement with experimental data.

\subsection{Proton decay problems}
\label{sec:ProtonDecay}

\subsubsection{The conventional approach to the proton decay problems}

Of course, these problems are well known and there are some standard solutions.
Let us specifically discuss the traditional cure of proton decay problems. The
$R$ parity violating terms can be forbidden by $R$ or matter parity
$\Z2^\mathcal{M}$ \cite{Farrar:1978xj,Dimopoulos:1981dw}, either of which is
usually assumed to be part of the definition of the MSSM. Formally, these two
symmetries differ by the transformation of the superpartners. However, there is
an intrinsic symmetry in any supersymmetric theory which sends the superspace
coordinate $\theta$ to minus itself.\label{ambiguity} Using this ambiguity, one
can easily convince one--self that the two symmetries are equivalent. After
imposing this symmetry, there are still the  dimension--five proton decay
operators, which can, however, be forbidden by baryon triality $B_3$
\cite{Ibanez:1991pr} (see \Tabref{tab:TraditionalSymmetries} for the charge
assignment).  
\begin{table}[h]
\begin{center}
  \begin{tabular}{|l||c|c|c|c|c|c|c|c|}
    \hline
   \phantom{$A^{A^A}$} & $Q$ & $\bar{U}$ & $\bar{D}$ & $L$ & $\bar{E}$ & $\hu$ & $\hd$ & $\bar{\nu}$ \\
    \hline\hline     
    $\Z2^\mathcal{M}$ 
          & $1$ & $1$  & $1$ &  $1$ & $1$ & $0$ & $0$ & $1$ \\
    $B_3$ & $0$ & $-1$ & $1$ & $-1$ & $2$ & $1$ & $-1$ & $0$ \\
    $P_6$ & $0$ & $1$  & $-1$ & $-2$ & $1$ & $-1$ & $1$ & $3$ \\
    \hline
  \end{tabular}
\end{center}
\caption{Matter parity $\Z2^\mathcal{M}$, baryon triality $B_3$ and proton
hexality $P_6$.}
\label{tab:TraditionalSymmetries}
\end{table}
The combination of both symmetries, i.e.\ $\Z2^\mathcal{M}$ times $B_3$, is
known as ``proton hexality'' $P_6$
\cite{Ibanez:1991pr,Babu:2003qh,Dreiner:2005rd}. The $P_6$ symmetry has several
very appealing features:
\begin{itemize}
 \item[\Large\Smiley] it forbids dimension--four and five proton decay
	operators;
 \item[\Large\Smiley] it allows the usual Yukawa couplings of the MSSM as well
 	as the Weinberg's neutrino operator $\kappa^{(0)}_{ij}\, \hu L_i\,\hu L_j$;
 \item[\Large\Smiley] it is the unique anomaly--free symmetry with the above
 features assuming traditional anomaly cancellation.
\end{itemize}
Unfortunately, $P_6$ has also some disturbing aspects:
\begin{itemize}
\item[\Large\Frowny] it is not consistent with unification of matter, i.e.\ it
 	is inconsistent with having universal discrete charges for all matter fields
 	(cf.\ \cite{Forste:2010pf});
\item[\Large\Frowny] it does not address the $\mu$ problem, i.e.\  it does not
 	provide us with a solution to all the above--mentioned problems of the MSSM.
\end{itemize} 

\subsection{Origin of proton decay operators in GUTs}

\begin{figure}[h]
\centerline{%
\subfigure[Dimension~6.]{\includegraphics{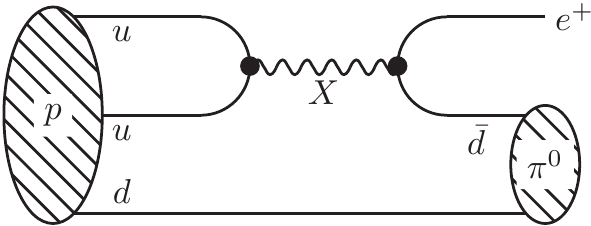}}
\qquad\qquad
\subfigure[Dimension~5.]{\includegraphics{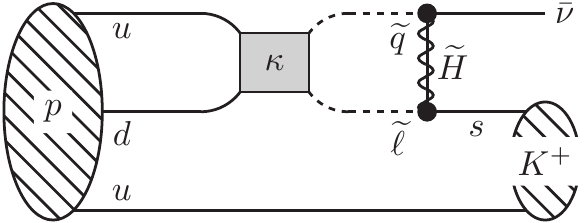}}
}
\caption{Dimension-- (a)~six and (b)~five proton decay diagrams in grand unified
theories, leading to the proton decay modes (a) $p\to\pi^0+e^+$ and (b) $p\to
K^++\bar\nu$. While the SUSY GUT prediction for (a) is still consistent with
experimental limits, the decay mode (b) often challenges explicit SUSY GUT
models.}
\label{fig:ProtonDecay}
\end{figure}

One may now wonder how serious the fact is that $P_6$ is not consistent with
(\SU5 or \SO{10}) unification. To this end, it is instructive to recall where
the proton decay operators come from.  One distinguishes between dimension 6 and
5 proton decay operators.  While the dimension--six operators can come from
gauge boson exchange (cf.\ \Figref{fig:ProtonDecay}~(a))~\cite{Georgi:1974sy},
the dimension--five ones (\Figref{fig:ProtonDecay}~(b)) may originate from the
color--triplet Higgs exchange \cite{Sakai:1981pk,Dimopoulos:1981dw}. While the
SUSY GUT predictions for the proton decay rates mediated by dimension--six
operators are still consistent with observation \cite{Murayama:2001ur}, the
dimension--five proton decay and the associated doublet--triplet splitting
problems cast some shadow on the scheme of (four--dimensional) SUSY GUT models
(cf.\ e.g.\ \cite{Dermisek:2000hr,Murayama:2001ur}). Some coefficients of the
$Q\,Q\,Q\,L$ operators have to be smaller than $10^{-8}/M_\mathrm{P}$
\cite{Hinchliffe:1992ad}, which leads to a lower bound on the color--triplet
Higgs mass far above $M_\mathrm{GUT}$ unless one arranges for very precise
cancellations between unrelated couplings (see e.g.\
\cite{EmmanuelCosta:2003pu,Bajc:2004xe}).

Given that $P_6$ is incompatible with grand unification, we see that this
symmetry cannot be used to solve the most severe problems of GUT models. This is
also in accordance with the fact that $P_6$ does not address the $\mu$ problem,
i.e.\ it cannot help us to understand the doublet--triplet splitting.

Various other solutions to the dimension--five proton decay problem of SUSY GUTs
rely on intricate GUT breaking sectors
\cite{Grinstein:1982um,Babu:1993we,Babu:2002fsa}. The Higgs fields typically
used for the GUT breaking and the generation of fermion masses are in
representations as large as \rep{75} of \SU{5} or \rep{126} of \SO{10}. The
corresponding large amount of $G_\mathrm{SM}$ charged states typically induces
large threshold corrections, which may clash with our basic assumption that
gauge unification is not an accident.

In what follows, we will therefore discuss alternative discrete symmetries which do not
suffer from these shortcomings. Specifically, we will identify anomaly--free
discrete symmetries which are consistent with (precision) gauge
unification and allow us to control the $\mu$ term.

\section{Non--anomalous discrete symmetries and unification}

In this section, we will first discuss (discrete) anomaly cancellation in
general. Then we will focus on symmetries that are consistent with unification
and forbid the $\mu$ term. In contrast to the traditional approach, we
make use of the Green--Schwarz (GS) mechanism for anomaly cancellation
\cite{Green:1984sg}.

\subsection{Anomaly universality}

We begin our discussion with the observation that, in the framework of grand unified
theories, once one allows for the Green--Schwarz mechanism, the requirement of
anomaly freedom is depleted to the demand of ``anomaly universality'', i.e.\
common anomaly coefficients of the SM gauge factors $G_i$.

Let us explain what that implies in practice. Consider, for example, the mixed
$G_i-G_i-\Z{N}$ anomaly coefficient for a $\Z{N}$ symmetry,
\begin{equation}
 A_{G_i-G_i-\Z{N}}~ =~ \sum_f \ell(\rep{r}^{(f)})\cdot q^{(f)}\;.
\end{equation} 
Here the sums extend over all fermion representations $\rep{r}^{(f)}$, while
$\ell^{(f)}$ denotes the Dynkin index of the fermions $f$ w.r.t.\ the gauge
group $G_i$ and $q^{(f)}$ are the discrete $\Z{N}$ charges. The traditional
anomaly constraints \cite{Ibanez:1991hv,Banks:1991xj}   correspond to the
condition that the $A_{G_i^2-\Z{N}}$ coefficients\footnote{Note that there are no
meaningful $\Z{N}^3$ anomaly constraints. This has been first shown in
\cite{Banks:1991xj} and can be seen more directly in the path integral approach
\cite{Araki:2008ek}.} have to vanish for all $G_i$,
\begin{equation}
 A_{G_i-G_i-\Z{N}}~=~0\mod\eta\quad \forall~G_i\;,
\end{equation}
where
\begin{equation}
 \eta~:=~\left\{\begin{array}{ll}
    N & \text{for $N$ odd}\;,\\
    N/2 & \text{for $N$ even}\;.
 \end{array}\right.
\end{equation}
On the other hand, ``anomaly universality'' only amounts to the requirement that
the anomaly coefficients be universal,
\begin{equation}
 A_{G_i-G_i-\Z{N}}~=~\rho\mod\eta\quad \forall~G_i\;,
\end{equation}
but that they do not necessarily have to vanish.
Here $\rho$ can be thought of as the contribution of a GS axion $a$, whose shift
transformation under the \Z{N} symmetry cancels the anomaly.
  
Where does the ``anomaly universality'' come from? Although universality of anomaly
coefficients is empirically found to be a property of most heterotic string
models \cite{Dine:2004dk,Araki:2008ek}, it is, as correctly pointed out in
\cite{Ludeling:2012cu}, in general not a necessary condition for anomaly
freedom. This can most easily be seen in the path integral formulation
\cite{Fujikawa:1979ay,Fujikawa:1980eg} of the GS mechanism (see e.g.\
\cite{Lee:2011dya,Chen:2012jg}). The crucial
ingredient is the coupling of the GS axion $a$ to the $F\,\widetilde{F}$ term of
the gauge group $G$. The GS axion $a$ is contained in the superfield $\dilaton$,
$\dilaton|_{\theta=0}=s+\I\,a$, and shifts under the symmetry transformation.
The GS anomaly cancellation requires the coupling
\begin{equation}\label{eq:DilatonCoupling}
 \int\!\D^2\theta\,f_S\, \dilaton\,W_\alpha W^\alpha
~\supset~\mathscr{L}\;.
\end{equation}
in the Lagrange density. Given this term, $s=\re\dilaton|_{\theta=0}$
contributes to $1/g^2$, see \cite{Lee:2011dya,Chen:2012jg} for more details. In
general, different couplings of $a$ to different SM gauge factors $G_i$
would allow for different $\rho$ constants for the different gauge
factors of the SM. However, in general, the ``saxion'' $s$ has a non--trivial
vacuum expectation value (VEV), such that non--universal couplings imply
non--universal contributions to $1/g_i^2$. This, in turn, would imply
that precision gauge unification is spoilt. Since this would
contradict our assumption that precision gauge unification is not an accident,
we will require anomaly universality in the rest of our discussion.       

\subsection{Non--\texorpdfstring{$\boldsymbol{R}$}{R} symmetries cannot forbid the
\texorpdfstring{$\boldsymbol{\mu}$}{mu} term in the MSSM}

Let us now look at discrete anomalies of non--$R$ symmetries in the MSSM. After
imposing \SU5 relations for the matter charges, the relevant anomaly
coefficients read
\begin{eqnarray}
 A_{\SU3^2-\Z{N}}
 & = &
 \frac{1}{2}\sum_{g=1}^3
 \left(3q_{\rep{10}}^g
 +q_{\crep{5}}^g\right)\;,\\
 A_{\SU2^2-\Z{N}}
 & = &
 \frac{1}{2}\sum_{g=1}^3
 \left(3q_{\rep{10}}^g
 +q_{\crep{5}}^g\right)
 +\frac{1}{2}\left(\qhu
 +\qhd\right)\;.
\end{eqnarray}
Here, in an obvious notation, $q_{\rep{10}}^g$ and $q_{\crep{5}}^g$ denote the
discrete charges of the $g^\mathrm{th}$ $\rep{10}$-- and $\crep{5}$--plet,
respectively, with $g$ playing the role of a generation index while $\qhu$ and
$\qhd$ are the charges of the Higgs doublets. Now, imposing anomaly universality,
i.e.\ demanding that
\begin{equation}
 A_{\SU2^2-\Z{N}}-A_{\SU3^2-\Z{N}}~=~0\mod\eta\;,
\end{equation} 
leads to a condition on the Higgs charges:
\begin{equation}
 \frac{1}{2}\left(q_{\hu}+q_{\hd}\right) 
 ~=~  0 \mod 
  \left\{\begin{array}{ll}
    N & \text{for $N$ odd}\;,\\
    N/2 & \text{for $N$ even}\;.
 \end{array}\right. 
\end{equation}
It is easy to see that this implies that the \Z{N} symmetry does not forbid the
Higgs bilinear. We hence see that ordinary, i.e.\  non--$R$, $\Z{N}$ symmetries
cannot forbid the $\mu$ term.

\subsection{Only discrete \texorpdfstring{$\boldsymbol{R}$}{R} symmetries may
forbid the \texorpdfstring{$\boldsymbol{\mu}$}{mu} term}

It is also obvious that, if anomaly--free discrete non--$R$ symmetries cannot
forbid the $\mu$ term, this also applies to continuous non--$R$ symmetries, for
which the anomaly constraints are even stronger. We are hence left with $R$
symmetries. Recalling that there are no anomaly--free continuous $R$ symmetries
in the MSSM \cite{Chamseddine:1995gb}, the only remaining option is the
discrete $R$ symmetries.

\subsection{\texorpdfstring{$\boldsymbol{R}$}{R} symmetries and 't Hooft anomaly matching}

't Hooft's concept of anomaly matching is a powerful tool for analyzing
symmetries \cite{'tHooft:1979bh}, which can also be used for discrete
symmetries \cite{Csaki:1997aw}. Let us spell this out for the case of discrete
$R$ symmetries in the MSSM, still assuming unification \cite{Chen:2012jg}.
Trivially, at the \SU5 level, there is only one anomaly coefficient,
\begin{equation} 
 A_{\SU5^2-\Z{M}^R}
 ~=~
 A_{\SU5^2-\Z{M}^R}^\mathrm{matter}
 +
 A_{\SU5^2-\Z{M}^R}^\mathrm{extra}
 +
 5q_\theta\;,
\end{equation}
which we have decomposed into the contribution from matter
$A_{\SU5^2-\Z{M}^R}^\mathrm{matter}$, extra states
$A_{\SU5^2-\Z{M}^R}^\mathrm{extra}$ and gauginos $5q_\theta$ with $q_\theta$
denoting the $R$ charge of the superspace
coordinate.\footnote{\label{ftn:qtheta}Note that there exists some confusion in
the literature. It is often assumed that the superpotential $\mathscr{W}$ has
$R$ charge 2, corresponding to $R$ charge 1 of the superspace coordinate,
$q_\theta=1$. However, as pointed out in \cite{Chen:2012jg}, one cannot, in
general, make this choice and, at the same time, demand that all discrete
charges are integer. We follow the convention that all discrete charges are
integer and keep $q_\theta$ variable.} $M$ is the order of the $R$ symmetry
transformation, which might be part of a larger symmetry. In addition to the
anomaly constraint from the whole gauge group, we can also consider the \SU3 and
\SU2 subgroups of \SU{5}. The corresponding anomaly coefficients read
\begin{subequations}
\begin{eqnarray}
 A^{\SU5}_{\SU3^2-\Z{M}^R}
 & = &
 A_{\SU3^2-\Z{M}^R}^\mathrm{matter}
 +
 A_{\SU3^2-\Z{M}^R}^\mathrm{extra}
 +
 3q_\theta
 +
 \frac{1}{2}\cdot2\cdot2\cdot q_\theta\;,
 \\
 A^{\SU5}_{\SU2^2-\Z{M}^R}
 & = &
 A_{\SU2^2-\Z{M}^R}^\mathrm{matter}
 +
 A_{\BLAU{\SU2}^2-\Z{M}^R}^\mathrm{extra}
 +
 2q_\theta
 +
 \frac{1}{2}\cdot2\cdot3\cdot q_\theta\;.
\end{eqnarray}
\end{subequations}
Here we have decomposed the gaugino contributions into their \SU3 and \SU2
parts, respectively, and into the contributions from $\SU5/G_\mathrm{SM}$. Assume
now that some mechanism eliminates the extra gauginos. This will lead to a
non--universality of the anomaly coefficients, which will, given our assumption that matter charges commute
with \SU5, have to be compensated for by the extra states. That is, the extra
states have to come in split multiplets. In other words,  't Hooft anomaly
matching for (discrete) $R$  symmetries implies the presence of split multiplets
below the GUT scale. The arguably simplest possibility to ``repair'' the gaugino
mismatch is to assume that there is a pair of massless weak doublets, which is
chiral w.r.t.\ $\Z{M}^R$, but no corresponding triplets. From this one infers
that, in the presence of an $R$ symmetry, the same mechanism that breaks the GUT
symmetry will also provide a mechanism for doublet--triplet splitting. However,
as we will discuss later, it is impossible to construct a four--dimensional
grand unified theory with a low energy $R$ symmetry without states beyond
those of the MSSM. This is also consistent with the observation that there are no
natural (in 't Hooft's sense) solutions to the doublet--triplet splitting
problem in such schemes.

\subsection{A unique discrete \texorpdfstring{$\boldsymbol{R}$}{R} symmetry for the MSSM}

Let us now impose, instead of \SU5 relations, stronger \SO{10} relations, i.e.\
that the charges $q$ for matter fields are universal. That is, consider a $\Z{M}^{R}$
symmetry under which quarks and leptons have the universal charge $q$. As we shall
demonstrate, this implies a unique symmetry \cite{Lee:2011dya,Chen:2012jg}. In
the first step, we require that the symmetry allows for $u$-- and $d$--type Yukawas,
implying that 
\begin{equation}
 2q + q_{\hu} ~ =~ 2q_\theta\mod M 
 \quad \text{and}\quad 
 2q + q_{\hd} ~ = ~ 2q_\theta\mod M\;.
\end{equation}
Subtracting these equations from each other,
\begin{equation}
 q_{\hu} - q_{\hd}~=~0\mod M\;,
\end{equation}
shows that also the charges of the two Higgs fields coincide.
The conditions for the presence of $u$--type Yukawa couplings and the Weinberg
operator are
\begin{equation}
 2q + q_{\hu} ~ =~ 2q_\theta \mod M 
 \quad \text{and} \quad
 2q + 2q_{\hu} ~ =~ 2q_\theta \mod M\;,
\end{equation} 
implying that $q_{\hu} =0\mod M$. Altogether we see that 
\begin{equation}
 q_{\hu} ~=~ q_{\hd} ~=~ 0\mod M
 \quad\text{and}\quad 
 q ~=~ q_\theta\mod M\;.
\end{equation}
From the conditions that the symmetry must be an $R$ symmetry,
\begin{equation}\label{eq:qnot0}
  q_\theta ~\neq~ 0 \mod \eta\;,
\end{equation}
and that it is ``anomaly universal'' in the MSSM,
\begin{equation}\label{eq:anomalyuniversal}
  A_{\SU3^2-\Z{M}^R} ~=~  3 q_\theta \mod \eta ~\stackrel{!}{=}~ q_\theta \mod \eta ~=~ A_{\SU2^2-\Z{M}^R}\;,
\end{equation}
it follows that $\eta$ is even which in turn implies that the order $M$ of the symmetry is a multiple of 4,
\begin{equation}
  M~=~4m, \quad m\in\mathbb{N}\;.
\end{equation}
Furthermore, given the ambiguity discussed on p.~\pageref{ambiguity},
equations~\eqref{eq:qnot0} and~\eqref{eq:anomalyuniversal} fix the $R$ charge of
the superspace coordinate $\theta$ to $q_\theta=m$. As a result, the simplest
non--trivial possibility is $M=4$ and $q=q_\theta=1$, i.e.\ a $\Z{4}^R$
symmetry. As is straightforward to see, the extensions to $\Z{4m}^R$ symmetries,
$m>1$, are trivial extensions as far as the MSSM is concerned. While it might
certainly be worthwhile to study such symmetries in the context of (singlet)
extensions of the MSSM, we can conclude that there is a unique symmetry for the
MSSM: a $\Z{4}^R$ with $q=q_\theta=1$ and $q_{\hu}=q_{\hd}=0$. 

This symmetry was first discussed in \cite{Babu:2002tx}. A version of the
uniqueness proof appeared in \cite{Lee:2010gv}. However, there it was assumed
that the superpotential has charge 2 in a normalization in which all discrete
charges are integer, which is, in general, not a valid assumption (cf.\ footnote
\ref{ftn:qtheta}). The uniqueness proof has been completed in
\cite{Chen:2012jg}.

The $\Z4^R$ anomaly coefficients are
\begin{subequations}
\begin{eqnarray}
 A_{\SU3^2-\Z{4}^R}                   
 & = & 
 6q-3q_\theta ~=~ q_\theta ~=~ 1 \mod4/2\;,\\
 A_{\SU2^2-\Z{4}^R}                   
 & = & 
 6q+\frac{1}{2}\left(q_{\hu}+q_{\hd}\right)
 -5q_\theta
 ~=~q_\theta ~=~ 1 \mod4/2\;,
\end{eqnarray}
\end{subequations}
The fact that the coefficients are non--trivial implies that the  $\Z4^R$ is
anomaly--free only via a non--trivial GS mechanism.

\subsection{GS anomaly cancellation and non--perturbative effects}

Let us briefly comment on the implications of GS anomaly cancellation. As
discussed above, the GS axion $a$ is contained in a superfield $\dilaton$,
$\dilaton|_{\theta=0}=s+\I\,a$. Since $a=\im\dilaton|_{\theta=0}$ shifts under
the $\Z{M}^R$ transformation, $R$ non--covariant superpotential terms can be
made invariant by multiplying them with $\mathrm{e}^{-b\,\dilaton}$. To be
specific, consider, as an example, the Higgs bilinear. The $\mu$ term is
obviously forbidden by the $\Z4^R$ symmetry, but the term
\begin{equation} 
 B\,\mathrm{e}^{-b\,\dilaton}\,\hu\hd
\end{equation}
will be allowed for appropriate values of $b$. In other words, the  holomorphic
$\mathrm{e}^{-b\,\dilaton}$ terms appear to violate the $\Z{M}^R$ symmetry. Such
terms have a well--known interpretation.  Given the coupling
\eqref{eq:DilatonCoupling}, $s=\re\dilaton|_{\theta=0}$ contributes to $1/g^2$,
and the holomorphic $B\,\mathrm{e}^{-b\,\dilaton}$ terms can be interpreted as
non--perturbative effects (cf.\ the ``retrofitting''
discussion~\cite{Dine:2006gm}). Altogether we see that there is a unique
symmetry of the MSSM that (i) forbids the $\mu$ term, (ii) is compatible with
\SO{10} and (iii) is anomaly--free; this symmetry has the feature  that the
$\mu$ term appears non--perturbatively and is  naturally suppressed.

\subsection{Further implications of \texorpdfstring{$\boldsymbol{\Z4^R}$}{Z4R}}

The $\Z4^R$ symmetry has important implications for the MSSM. Among the gauge
invariant terms shown in \eqref{eq:WMSSM}, the $\mu$ term, the $R$ parity
violating terms and the dimension five proton decay operators are forbidden at
the perturbative level while, by construction, the Yukawa couplings and the
Weinberg operator are allowed. As discussed above, $\mu$ and the dimension--five
proton decay operators appear at the non--perturbative level, whereas the $R$
parity violating terms are still forbidden at the non--perturbative level by a
``non--anomalous'' \Z2 subgroup  which is equivalent to matter parity. How can
one determine the size of the non--perturbative terms? The order parameter for
$R$ symmetry breaking is the superpotential VEV $\langle\mathscr{W}\rangle$, or,
in other words, the gravitino mass $m_{\nicefrac{3}{2}}$. Hence
\begin{equation}
 \mu~\sim~m_{\nicefrac{3}{2}}~\simeq~\langle\mathscr{W}\rangle/\MP^2
\end{equation}
with $\MP$ denoting the Planck scale. The non--perturbatively generated
dimension--five proton decay operators are phenomenologically harmless, 
\begin{equation}
 \kappa^{(1,2)}_{ijk\ell}~\sim~
 m_{\nicefrac{3}{2}}/\MP^2~\ll~ 10^{-8}/\MP\;,
\end{equation}
where we compare the theoretical expectation with the experimental constraints
\cite{Hinchliffe:1992ad}.

\section{No \texorpdfstring{$\boldsymbol{R}$}{R} symmetries in conventional 4D GUTs}

In the previous section, we have seen that only $R$ symmetries can forbid the $\mu$ term in the MSSM.
However, as we shall show now, $R$ symmetries are not available in
four--dimensional GUTs \cite{Fallbacher:2011xg}. More specifically, if one assumes 
\begin{itemize}
 \item[(i)] a GUT model in four dimensions based on $G\supset\SU5$,
 \item[(ii)] that the GUT symmetry breaking is spontaneous, and
 \item[(iii)] that there is only a finite number of fields,
\end{itemize}
one can prove that it is impossible to get a low--energy
effective theory with both
\begin{itemize}
 \item[1.] just the MSSM field content, and
 \item[2.] residual $R$ symmetries.
\end{itemize}
For the purposes of this review, we will restrict ourselves to presenting the basic
argument. Consider an \SU5 model with an (arbitrary) $R$ symmetry and 
a chiral $\rep{24}$--plet
breaking $\SU5\to G_\mathrm{SM}$. Recall the branching rule
\begin{equation}
 \rep{24}~\to~
 (\rep{8},\rep{1})_0
 \oplus
 (\rep{1},\rep{3})_0
 \oplus
 (\rep{3},\rep{2})_{-\nicefrac{5}{6}}
 \oplus
 (\crep{3},\rep{2})_{\nicefrac{5}{6}}
 \oplus
 (\rep{1},\rep{1})_{0}
 \;.
\end{equation}
Since the \rep{24}--plet attains a VEV but may not break the $R$ symmetry, it
has to have $R$ charge 0.  In the course of GUT breaking, the multiplets
$(\rep{3},\rep{2})_{-\nicefrac{5}{6}}\oplus(\crep{3},\rep{2})_{\nicefrac{5}{6}}$
are absorbed by the extra gauge bosons from $\SU5/G_\mathrm{SM}$. Thus, there
are extra massless states in the representations
$(\rep{8},\rep{1})_0\oplus(\rep{1},\rep{3})_0$, whose masses are forbidden by
the $R$ symmetry.

One can now ask the question whether it is possible to make these unwanted
states massive. It is easy to see that the introduction of extra
$\rep{24}$--plets with $R$ charge 2 only shifts the problem of massless states
to different representations. In particular, in this case there would be
massless states in the representation $(\rep{3},\rep{2})_{-\nicefrac{5}{6}}
\oplus (\crep{3},\rep{2})_{\nicefrac{5}{6}}$ representations. Repeating this
argument inductively shows that with a  finite number of $\rep{24}$--plets one
will always have massless exotics.  The only way to circumvent this argument  
is to have infinitely many $\rep{24}$--plets.

It is possible to generalize the basic argument to 
\begin{itemize}
 \item arbitrary \SU5 representations;
 \item larger GUT groups $G\supset\SU5$; 
 \item singlet extensions of the MSSM.
\end{itemize}
The proof can be found in \cite{Fallbacher:2011xg}. Here we shall only discuss the
implications of these statements. A `natural' solution of the $\mu$ and/or
doublet--triplet splitting problem requires a symmetry that forbids $\mu$. So
far we have learned that:
\begin{enumerate} 
\item only $R$ symmetries can forbid the $\mu$ term;
\item anomaly matching requires the existence of split multiplets;
\item $R$ symmetries are not available in 4D GUTs.
\end{enumerate}
This implies that `natural' solutions to the  $\mu$ and/or doublet--triplet
splitting problems   are not available in four dimensions! This might be
interpreted as the necessity to go to models with extra dimensions, such as string
compactifications.

\section{Higher--dimensional and string models}

In this section, we will discuss how going to extra dimensions allows us to
evade the no--go theorem presented in the previous section. In such settings it
is moreover possible to answer the question of the origin of $R$ symmetries and
one has better control over the higher--dimensional operators such as the
effective $\mu$ term.

\subsection{Grand unification in higher dimensions}

It is often stated that higher--dimensional GUTs appear more ``appealing''. This
is because new possibilities of symmetry breaking arise
\cite{Witten:1985xc,Breit:1985ud}. In addition, the Kaluza--Klein towers
provide us with, from a four--dimensional point of view, infinitely many states 
(cf.\ the discussion in \cite{Goodman:1985bw}), thus allowing us to to evade the
no--go theorem.

What is more, $R$ symmetries have a clear geometric interpretation. They originate
from the Lorentz symmetry of compact dimensions (cf.\ e.g.\ the discussion in
\cite{Nilles:2012cy}) and are arguably on the same footing as the fundamental
symmetries $C$, $P$ and $T$.

\subsection{Extra--dimensional/Stringy origin of \texorpdfstring{$\boldsymbol{\Z4^R}$}{Z4R}}

String models offer a geometric explanation of discrete symmetries (for a recent
review see e.g.~\cite{Nilles:2012cy}). Specifically, in stringy heterotic
orbifolds, one obtains effective theories with residual discrete $R$ symmetries.
In particular, one can determine the $R$ charges of the different states. Such
models often exhibit a $\Z4^R$ symmetry, under which localized fields have odd $R$
charges while bulk fields have even $R$ charges. This harmonizes nicely with the
scheme of ``local grand unification'' \cite{Buchmuller:2005sh} where matter
fields are localized in regions with \SO{10} symmetry and, therefore, come in
complete \SO{10} multiplets, while Higgs fields come from the bulk and, therefore,
are split.\footnote{In concrete models the third family comes partially
from the bulk \cite{Lebedev:2007hv} (and is a so--called ``patchwork family''
\cite{Pena:2012ki}, among other things giving rise to gauge--top unification
\cite{Hosteins:2009xk}).}

Let us now discuss globally consistent string models with these features
\cite{Blaszczyk:2009in,Kappl:2010yu}. These are  $\Z2\times\Z2$ orbifold models
with the exact MSSM spectrum. They exhibit vacua, i.e.\  field  configurations
that preserve supersymmetry perturbatively, with various good features
\begin{itemize}
 \item[\checkmark] non--local GUT breaking; 
 \item[\checkmark] no `fractionally charged exotics'; 
 \item[\checkmark] (most) exotics decouple at the linear level in SM singlets, i.e.\ just MSSM
 below GUT scale with masslessness of Higgs fields ensured by $\Z4^R$;
 \item[\checkmark] non--trivial full--rank Yukawa couplings;
 \item[\checkmark] gauge--top unification;
 \item[\checkmark] SU(5) relation $y_\tau\simeq y_b$.
\end{itemize}
Note that these are, unfortunately, just toy models since they exhibit certain
unrealistic features such as \SU5 Yukawa relations also for light generations.
Nevertheless such models illustrate that a successful string embedding of the $
\Z4^R$ symmetry is possible.

\section{Conclusions}

In this review, we have discussed the role of $R$ symmetries in supersymmetric
models that give rise to (precision) gauge unification. Specifically, we have
made the following assumptions:
\begin{itemize}
 \item[(i)] anomaly freedom (allowing for GS anomaly cancellation);
 \item[(ii)] $\mu$ term forbidden at the perturbative level;
 \item[(iii)] Yukawa couplings and Weinberg neutrino mass operator allowed;
 \item[(iv)] \SU5 or \SO{10} GUT relations for quarks and leptons.
\end{itemize}
We have then shown that
\begin{itemize}
 \item[1.] assuming (i) and \SU5 relations, only $R$ symmetries can forbid the $\mu$ term
 in the MSSM;
 \item[2.] assuming (i)--(iii) and \SO{10} relations, there is a unique $\Z4^R$ symmetry;
 \item[3.] $R$ symmetries are not available in 4D GUTs, implying that there is 
 no `natural' solution to doublet--triplet splitting in four dimensions.
\end{itemize}
The simple anomaly--free $\Z4^R$ symmetry turns out to provide a solution to the
$\mu$ problem and, as a bonus, automatically suppresses proton decay operators.
Models with this symmetry predict that proton decay proceeds via dimension six
operators, i.e.\ via gauge boson exchange. Yet, since such settings cannot be
embedded into four--dimensional GUTs, one will have to analyze
higher--dimensional models in order to make more detailed predictions.

\begin{figure}[h]
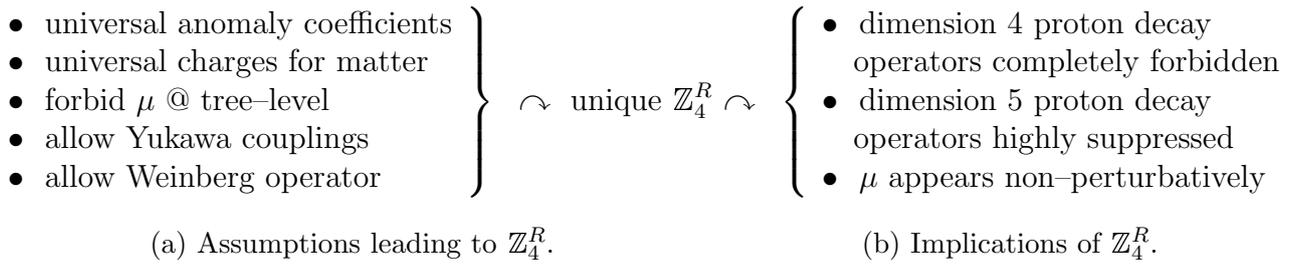

\centerline{%
\subfigure[Assumptions leading to $\Z4^R$.]{$\displaystyle
\left.\begin{array}{l}
\text{\textbullet~~universal anomaly coefficients}\\
\text{\textbullet~~universal charges for matter}\\
\text{\textbullet~~forbid $\mu$ @ tree--level}\\
\text{\textbullet~~allow Yukawa couplings}\\
\text{\textbullet~~allow Weinberg operator}
\end{array}\right\}
~\curvearrowright~\text{unique}~\Z4^R$}
\subfigure[Implications of $\Z4^R$.]{$\displaystyle
\curvearrowright~
\left\{\begin{array}{l}
\text{\textbullet~~dimension 4 proton decay}\\~~~\text{operators completely forbidden}\\
\text{\textbullet~~dimension 5 proton decay}\\~~~\text{operators highly suppressed}\\
\text{\textbullet~~$\mu$ appears non--perturbatively}\\
\end{array}\right.
$}}
\caption{(a) Assumptions leading to the unique $\Z4^R$ and (b) implications.}
\end{figure}

Deriving the $\Z4^R$ symmetry from string theory allows us to understand where
it comes from: it can arise as a discrete remnant of Lorentz symmetry in extra
dimensions. Guided by this $\Z4^R$ symmetry we have reported on a globally
consistent string model with (i) the exact MSSM spectrum;  (ii)
non--local/Wilson line GUT breaking; (iii) non--trivial full--rank Yukawa
couplings; (iv) exact matter parity; (v) $\mu\sim m_{\nicefrac{3}{2}}$ and (vi)
dimension--five proton decay operators sufficiently suppressed.

\section*{Acknowledgments}

M.-C.C.\ would like to thank TU M\"unchen, where part of the work  was done, for
hospitality. M.R.\ would like to thank the  UC Irvine, where part of this work
was done, for  hospitality. This work was partially supported by the DFG
cluster  of excellence ``Origin and Structure of the Universe'' and the
Graduiertenkolleg ``Particle Physics at the Energy Frontier of New Phenomena''
by Deutsche Forschungsgemeinschaft (DFG). The work of M.-C.C.\ was supported, in
part, by the U.S.\ National Science  Foundation under Grant No.\ PHY-0970173. 
M.-C.C.\ and  M.R.\  would like to thank CETUP* for  hospitality and support.
This research was done in the context of the ERC  Advanced Grant project
``FLAVOUR''~(267104).

\bibliography{Orbifold}

\providecommand{\bysame}{\leavevmode\hbox to3em{\hrulefill}\thinspace}
\frenchspacing
\newcommand{\origttfamily}{}
\let\origttfamily=\ttfamily
\renewcommand{\ttfamily}{\origttfamily \hyphenchar\font=`\-}

\begin{thebibliography}{10}

\bibitem{Dimopoulos:1981yj}
S.~Dimopoulos, S.~Raby, and F.~Wilczek, Phys. Rev. \textbf{D24} (1981), 1681.

\bibitem{Raby:2008gh}
S.~Raby, Eur. Phys. J. \textbf{C59} (2009), 223, \texttt{arXiv:0807.4921}
  [hep-ph].

\bibitem{Georgi:1974sy}
H.~Georgi and S.~L. Glashow, Phys. Rev. Lett. \textbf{32} (1974), 438.

\bibitem{Fritzsch:1974nn}
H.~Fritzsch and P.~Minkowski, Ann. Phys. \textbf{93} (1975), 193.

\bibitem{Allanach:1999ic}
B.~C. Allanach, A.~Dedes, and H.~K. Dreiner, Phys. Rev. \textbf{D60} (1999),
  075014, \texttt{hep-ph/9906209}.

\bibitem{Farrar:1978xj}
G.~R. Farrar and P.~Fayet, Phys. Lett. \textbf{B76} (1978), 575.

\bibitem{Dimopoulos:1981dw}
S.~Dimopoulos, S.~Raby, and F.~Wilczek, Phys. Lett. \textbf{B112} (1982), 133.

\bibitem{Ibanez:1991pr}
L.~E. Ib{\'a}{\~n}ez and G.~G. Ross, Nucl. Phys. \textbf{B368} (1992), 3.

\bibitem{Babu:2003qh}
K.~S. Babu, I.~Gogoladze, and K.~Wang, Phys. Lett. \textbf{B570} (2003), 32,
  \texttt{hep-ph/0306003}.

\bibitem{Dreiner:2005rd}
H.~K. Dreiner, C.~Luhn, and M.~Thormeier, Phys. Rev. \textbf{D73} (2006),
  075007, \texttt{hep-ph/0512163}.

\bibitem{Forste:2010pf}
S.~F{\"o}rste, H.~P. Nilles, S.~Ramos-S{\'a}nchez, and P.~K.~S. Vaudrevange,
  \texttt{arXiv:1007.3915} [hep-ph].

\bibitem{Sakai:1981pk}
N.~Sakai and T.~Yanagida, Nucl. Phys. \textbf{B197} (1982), 533.

\bibitem{Murayama:2001ur}
H.~Murayama and A.~Pierce, Phys.Rev. \textbf{D65} (2002), 055009,
  \texttt{arXiv:hep-ph/0108104} [hep-ph].

\bibitem{Dermisek:2000hr}
R.~Derm\'{i}\u{s}ek, A.~Mafi, and S.~Raby, Phys. Rev. \textbf{D63} (2001),
  035001, \texttt{hep-ph/0007213}.

\bibitem{Hinchliffe:1992ad}
I.~Hinchliffe and T.~Kaeding, Phys. Rev. \textbf{D47} (1993), 279.

\bibitem{EmmanuelCosta:2003pu}
D.~Emmanuel-Costa and S.~Wiesenfeldt, Nucl.Phys. \textbf{B661} (2003), 62,
  \texttt{arXiv:hep-ph/0302272} [hep-ph].

\bibitem{Bajc:2004xe}
B.~Bajc, A.~Melfo, G.~Senjanovi{\'c}, and F.~Vissani, Phys. Rev. \textbf{D70}
  (2004), 035007, \texttt{hep-ph/0402122}.

\bibitem{Grinstein:1982um}
B.~Grinstein, Nucl.Phys. \textbf{B206} (1982), 387.

\bibitem{Babu:1993we}
K.~S. Babu and S.~M. Barr, Phys. Rev. \textbf{D48} (1993), 5354,
  \texttt{hep-ph/9306242}.

\bibitem{Babu:2002fsa}
K.~Babu and S.~Barr, Phys.Rev. \textbf{D65} (2002), 095009,
  \texttt{arXiv:hep-ph/0201130} [hep-ph].

\bibitem{Green:1984sg}
M.~B. Green and J.~H. Schwarz, Phys. Lett. \textbf{B149} (1984), 117.

\bibitem{Ibanez:1991hv}
L.~E. Ib{\'a}{\~n}ez and G.~G. Ross, Phys. Lett. \textbf{B260} (1991), 291.

\bibitem{Banks:1991xj}
T.~Banks and M.~Dine, Phys. Rev. \textbf{D45} (1992), 1424,
  \texttt{hep-th/9109045}.

\bibitem{Araki:2008ek}
T.~Araki et~al., Nucl. Phys. \textbf{B805} (2008), 124,
  \texttt{arXiv:0805.0207} [hep-th].

\bibitem{Dine:2004dk}
M.~Dine and M.~Graesser, JHEP \textbf{01} (2005), 038, \texttt{hep-th/0409209}.

\bibitem{Ludeling:2012cu}
C.~L{\"u}deling, F.~Ruehle, and C.~Wieck, Phys.Rev. \textbf{D85} (2012),
  106010, \texttt{arXiv:1203.5789} [hep-th].

\bibitem{Fujikawa:1979ay}
K.~Fujikawa, Phys. Rev. Lett. \textbf{42} (1979), 1195.

\bibitem{Fujikawa:1980eg}
K.~Fujikawa, Phys. Rev. \textbf{D21} (1980), 2848.

\bibitem{Lee:2011dya}
H.~M. Lee, S.~Raby, M.~Ratz, G.~G. Ross, R.~Schieren, K.~Schmidt-Hoberg, and
  P.~K. Vaudrevange, Nucl.Phys. \textbf{B850} (2011), 1,
  \texttt{arXiv:1102.3595} [hep-ph].

\bibitem{Chen:2012jg}
M.-C. Chen, M.~Ratz, C.~Staudt, and P.~K. Vaudrevange, Nucl.Phys. \textbf{B866}
  (2012), 157, \texttt{arXiv:1206.5375} [hep-ph].

\bibitem{Chamseddine:1995gb}
A.~H. Chamseddine and H.~K. Dreiner, Nucl.Phys. \textbf{B458} (1996), 65,
  \texttt{arXiv:hep-ph/9504337} [hep-ph].

\bibitem{'tHooft:1979bh}
G.~'t~Hooft, NATO Adv. Study Inst. Ser. B Phys. \textbf{59} (1980), 135.

\bibitem{Csaki:1997aw}
C.~Cs{\'{a}}ki and H.~Murayama, Nucl. Phys. \textbf{B515} (1998), 114,
  \texttt{hep-th/9710105}.

\bibitem{Babu:2002tx}
K.~S. Babu, I.~Gogoladze, and K.~Wang, Nucl. Phys. \textbf{B660} (2003), 322,
  \texttt{hep-ph/0212245}.

\bibitem{Lee:2010gv}
H.~M. Lee, S.~Raby, M.~Ratz, G.~G. Ross, R.~Schieren, K.~Schmidt-Hoberg, and
  P.~K. Vaudrevange, Phys.Lett. \textbf{B694} (2011), 491,
  \texttt{arXiv:1009.0905} [hep-ph].

\bibitem{Dine:2006gm}
M.~Dine, J.~L. Feng, and E.~Silverstein, Phys. Rev. \textbf{D74} (2006),
  095012, \texttt{hep-th/0608159}.

\bibitem{Fallbacher:2011xg}
M.~Fallbacher, M.~Ratz, and P.~K. Vaudrevange, Phys.Lett. \textbf{B705} (2011),
  503, \texttt{arXiv:1109.4797} [hep-ph].

\bibitem{Witten:1985xc}
E.~Witten, Nucl. Phys. \textbf{B258} (1985), 75.

\bibitem{Breit:1985ud}
J.~D. Breit, B.~A. Ovrut, and G.~C. Segre, Phys. Lett. \textbf{B158} (1985),
  33.

\bibitem{Goodman:1985bw}
M.~W. Goodman and E.~Witten, Nucl.Phys. \textbf{B271} (1986), 21.

\bibitem{Nilles:2012cy}
H.~P. Nilles, M.~Ratz, and P.~K. Vaudrevange, \texttt{arXiv:1204.2206}
  [hep-ph].

\bibitem{Buchmuller:2005sh}
W.~Buchm{\"u}ller, K.~Hamaguchi, O.~Lebedev, and M.~Ratz,
  \texttt{arXiv:hep-ph/0512326} [hep-ph], 143.

\bibitem{Lebedev:2007hv}
O.~Lebedev, H.~P. Nilles, S.~Raby, S.~Ramos-S{\'a}nchez, M.~Ratz, P.~K.~S.
  Vaudrevange, and A.~Wingerter, Phys. Rev. \textbf{D77} (2007), 046013,
  \texttt{arXiv:0708.2691 [hep-th]}.

\bibitem{Pena:2012ki}
D.~K.~M. Pena, H.~P. Nilles, and P.-K. Oehlmann, \texttt{arXiv:1209.6041}
  [hep-th].

\bibitem{Hosteins:2009xk}
P.~Hosteins, R.~Kappl, M.~Ratz, and K.~Schmidt-Hoberg, JHEP \textbf{07} (2009),
  029, \texttt{arXiv:0905.3323} [hep-ph].

\bibitem{Blaszczyk:2009in}
M.~Blaszczyk, S.~G. Nibbelink, M.~Ratz, F.~Ruehle, M.~Trapletti, et~al.,
  Phys.Lett. \textbf{B683} (2010), 340, \texttt{arXiv:0911.4905} [hep-th].

\bibitem{Kappl:2010yu}
R.~Kappl, B.~Petersen, S.~Raby, M.~Ratz, R.~Schieren, and P.~K. Vaudrevange,
  Nucl.Phys. \textbf{B847} (2011), 325, \texttt{arXiv:1012.4574} [hep-th].

\end{thebibliography}
\addcontentsline{toc}{section}{Bibliography}
\bibliographystyle{NewArXiv} 
\end{document}